\documentclass[useAMS,usenatbib]{mn2e}
\usepackage{graphicx}
\begin{document}
\title[The 1997 event in the Crab pulsar revisited]{The 1997 event in the Crab pulsar revisited}
\author[F.Graham Smith, A.G.Lyne, and C.Jordan]{ F.Graham Smith, A.G.Lyne, and C.Jordan\\
Jodrell Bank Centre for Astrophysics\\
The Alan Turing Building\\
School of Physics and Astronomy\\
The University of Manchester\\
Oxford Rd\\
Manchester\\
M13 9PL\\
UK\\}
\maketitle
\parskip=10pt

\begin{abstract}
A complex event observed in the radio pulses from the Crab pulsar in
1997 included echoes, a dispersive delay, and large changes in
intensity.  It is shown that these phenomena were due to refraction at
the edge of a plasma cloud in the outer region of the Crab Nebula.  Several similar events have been 
observed, although in less detail.  It is suggested that the plasma cloud is in the form of filaments 
with diameter around $3\times 10^{11}$ m and electron density of order $10^4$ cm$^{-3}$.
\end{abstract}


\section{Introduction}

A dramatic event in the radio pulses received from the Crab Pulsar was
observed in October 1997, during routine observations at Green Bank and
at Jodrell Bank (Backer et al. 2000; Smith and Lyne
2000)\nocite{bwv00}\nocite{sl00}. A prominent part of the event was
an echo, following both the main pulse and the interpulse, which was
attributed to multi-path propagation involving an ionised cloud at a
large distance from the pulsar but within the Crab Nebula. As the
cloud had crossed the line of sight, a large reduction of intensity was observed, followed by an increase in dispersion measure.
Backer et al. describe the event in detail.  They propose a
model describing the cloud as a prism, attributing the echo to a ray
path through the main body of the prism, and the reduction in
intensity to a lensing effect as the thick base of the prism crossed
the line of sight. Our paper is a revision of this model.

The echo observed on this occasion, and on several others (Lyne et
al. 2001\nocite{lpg01}), had two components, the first with reducing
delay followed after several weeks by a second with increasing delay. It now seems that a
more appropriate model would be an essentially symmetric model,
involving refraction at the two edges of a cloud rather than within
the main body of the cloud.  The lensing effect is then the cause of
both the delayed ray path and the reduction in intensity.  We suggest
that the cloud is in the form of a filament, which crosses the line of
sight due to the known proper motion of the pulsar.

A similar analysis has been applied by 
Crossley et al (2007)\nocite{ceh07} to echoes of giant pulses from the Crab pulsar; these echoes had much shorter delays, and were attributed to structure in the inner region of the Crab nebula. 

\begin{figure*}
\centering
\includegraphics [width=8cm]{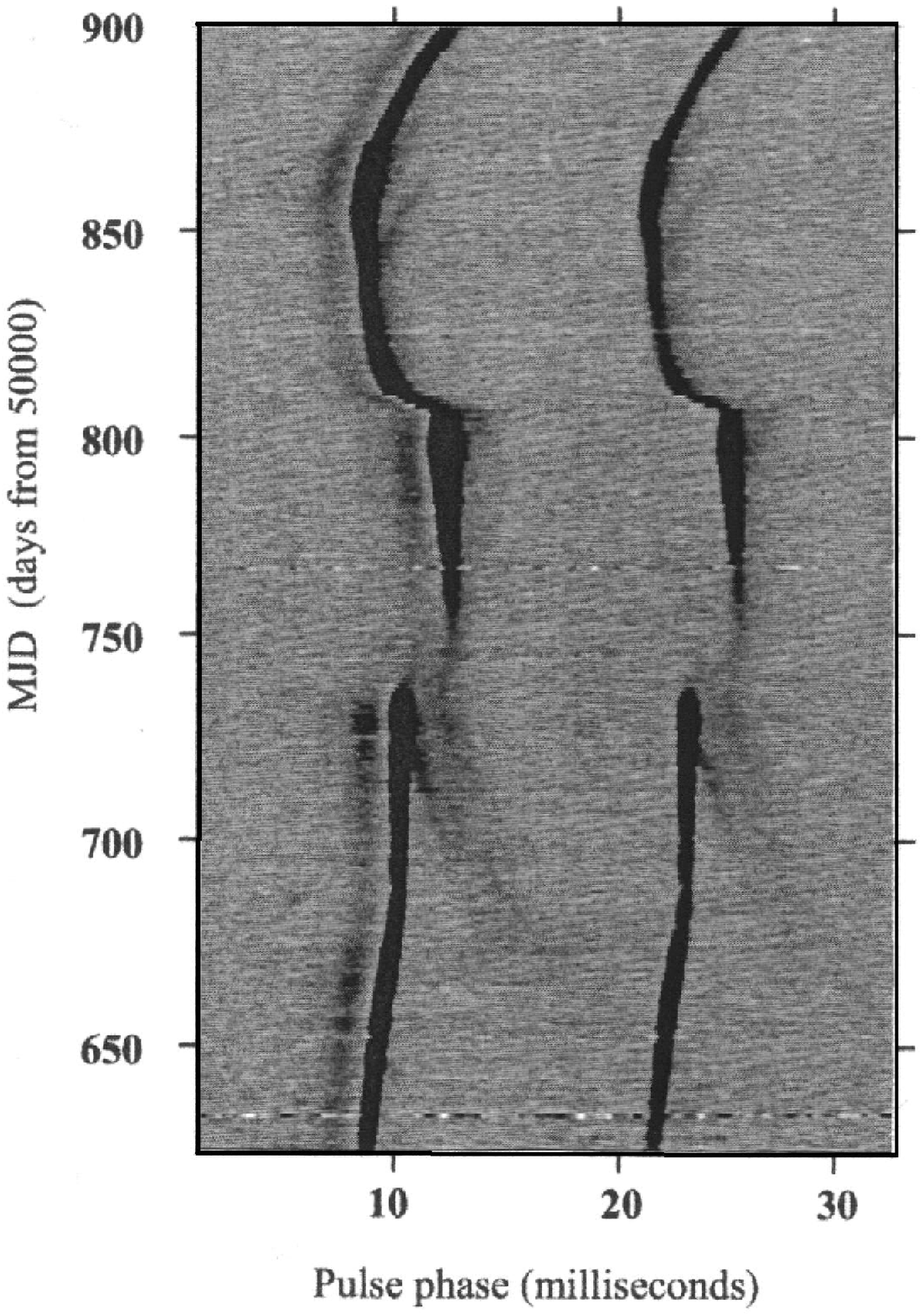}
\includegraphics [width=8cm]{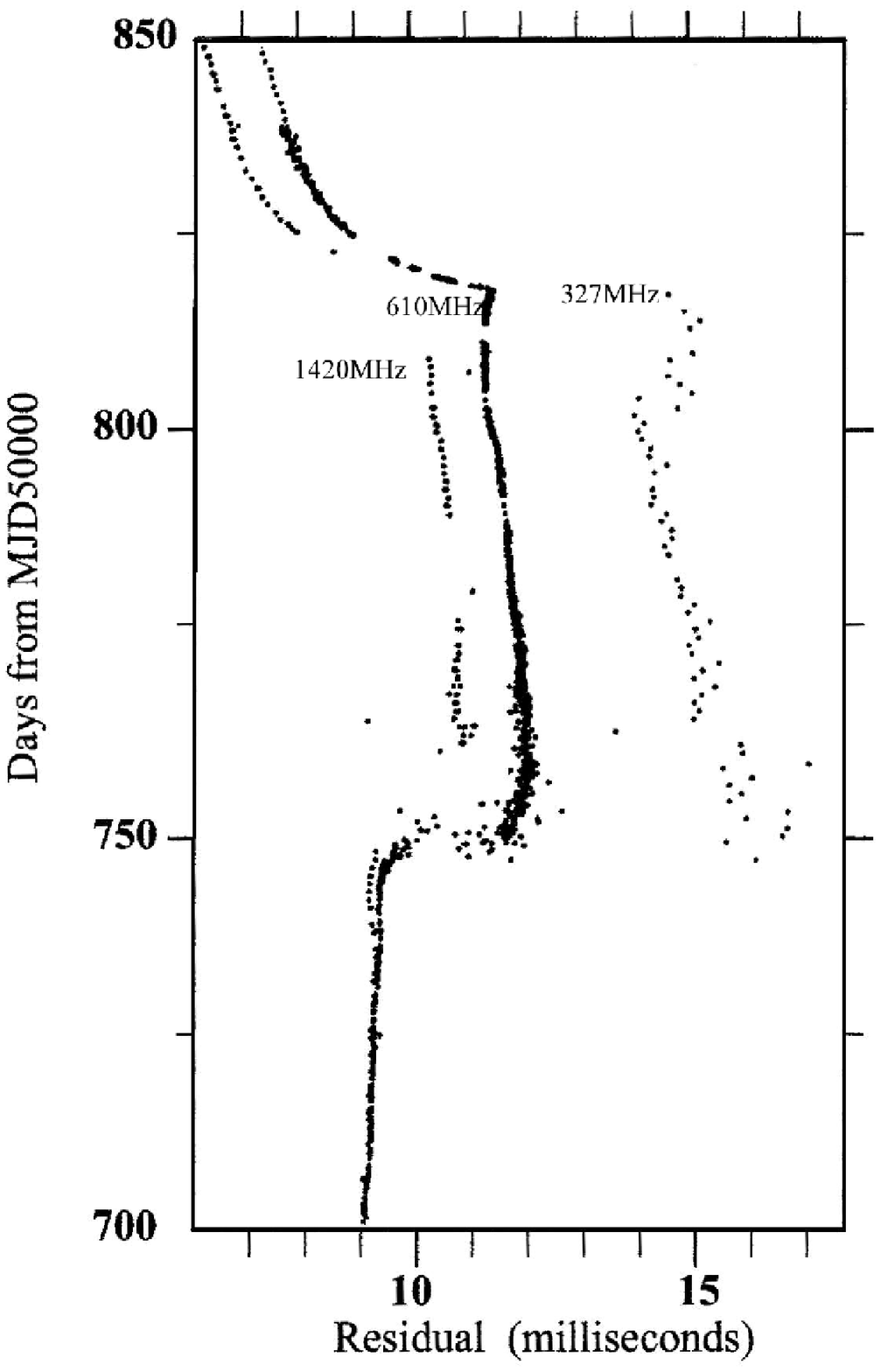}
\caption{(a) The pulses of the Crab pulsar, recorded at 610MHz by the 13-m
radio telescope at Jodrell Bank between 1997 July 17 and December 29
(MJD 50646-50811). Each observation is typically an integration over
12 h, obtained by adding the intensity synchronously at the pulse period
(approximately 33 ms). Each horizontal scan shows the
three pulse components, which normally trace a vertical line when the
pulsar rotation follows a constant slowdown rate. The echoes are seen as diffuse emission following both
the main pulse and the interpulse. (b) Pulse arrival times at 1420, 610 and 327 MHz.}
\label{picture}
\end{figure*}

\section{The observed phenomena}
  
  The sequence of events is shown in Figure ~\ref{picture}a, which shows
the evolution of the pulse profile as observed at 610 MHz during 260
days starting at MJD 50620.
  Data from the daily monitoring of the Crab pulsar (Lyne et al.
1993) have been used to provide a single high signal-to-noise
profile of the pulse each day. Each observation is typically an integration over
12 h, obtained by adding the intensity synchronously at the pulse period
(approximately 33 ms). The pulse profile is presented as a grey-scale of
intensity on a vertical line in the figure (this is the only available format 
for this observation). The full rotation period of 33
milliseconds is displayed, and a constant value of slowdown rate
$\dot\nu$ is used throughout. 

The time-scale is adjusted to an ephemeris determined from the pulse arrival times over a period of one
month near the middle of the sequence; a vertical track of the pulses in the
figure would indicate rotation with normal slowdown. The main pulse is
preceded by the much weaker precursor and followed 0.013 s later
by the interpulse.   The behaviour from day 620 to 670 is
normal.  From day 670 to 720 a faint echo is seen following the two most intense
components, with a delay decreasing from about 5 ms to zero at or near
day 720. (We have some evidence that a similar echo from the precursor
was present, but it is obscured by the more intense main pulse.)  The
intensity of the echo increases from about 2$\%$ to 5$\%$ of the normal
pulse as the delay decreases.

The intensity of the pulse is known to vary due to interstellar
scintillation, but the increase in intensity from day 730 to 740 is
unusual and significant.  At day 740 all three normal components fade
and disappear; they are replaced on day 747 by the same pattern with a
delay of 2.3 ms and with intensity increasing until about day 800.  On
day 812 a glitch occurred; this is a speed-up following the pattern of
many previous glitches, and it was presumably unrelated to the other
phenomena.  The increasing delay after day 860 indicates an increase
in slowdown rate, as is normally observed after a glitch in this
pulsar.  At day 845 an echo is again observed, similar to the earlier
echo but with increasing rather than decreasing delay. A similar faint
receding echo is also seen at day 800.

Observations were also made at frequencies of 327 MHz (Backer et al 2000) and 1420 MHz (Jodrell Bank). 
These show 
 similar behaviour, but with different step delays.  Figure \ref{picture}b shows the delays as timing residuals over a period of 150 days which includes all the main events.  The pulse arrival times are adjusted to coincide, with zero slope, during the first 30 days. The delay is seen
to be dispersive, and previous interpretations have attributed it to
propagation through an ionised cloud within the Crab Nebula but at a
large distance from the pulsar. It was remarked that
the delay did not follow precisely the normal dispersion law, in which the delay
should be proportional to $\nu^{-2}$.

\begin{figure*}
\centering
\resizebox{120mm}{!}{\includegraphics{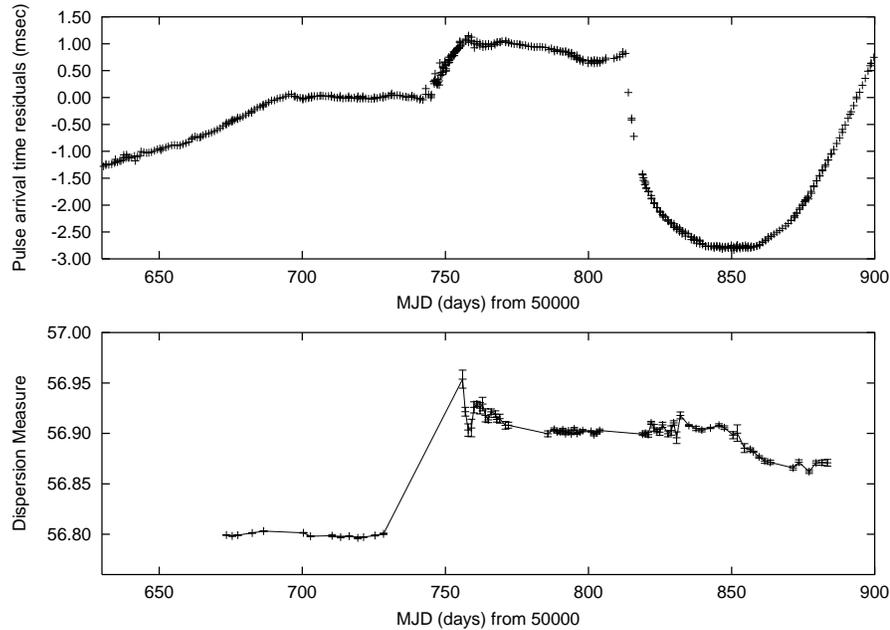}}
\vspace {4cm}
\caption{a Pulse arrival times corrected for dispersion. b. The dispersion measure (DM) observed over 330 days. }
\label{dispersion}
\end{figure*}
The timing observations shown in Figure \ref{picture}b were used to obtain values of dispersion measure over a period of 330 days.  Figure \ref{dispersion}a shows the pulse arrival times corrected for dispersion, \ref{dispersion}b shows the dispersion measure.  Before the event,
the dispersion measure was substantially constant at 56.80 cm$^{-3}$pc; it increases
abruptly to 56.92 cm$^{-3}$pc when the pulse reappears on day 750. It
then decreases almost monotonically over the following 250 days to
56.87 cm$^{-3}$pc. (Note particularly the steeper fall around day 850,
the time of the departing echo.) The value was unaffected by the
glitch at day 812.  It is this episode of increased dispersion that has
naturally been interpreted as the passage of a plasma cloud across the
line of sight to the pulsar.

\section{The echo delay}

The echo delay follows the same parabolic path at both 327 MHz and 610
MHz, from a maximum of about 5 milliseconds down to zero; the
departing echo observed at day 845 follows a mirror image of the same
geometry.  This indicates a maximum extra geometrical path $\delta =
1.5\times10^3$ km.  The echo is evidently due to an object which
approaches and moves across the line of sight.  In previous analyses,
Backer et al. (2000) and Lyne et al. (2001) considered the echo as a specular reflection from the surface of a
cloud; we now consider it in terms of refraction within the cloud.

\section{A plasma lens}

\begin{figure}
\includegraphics[width=8cm]{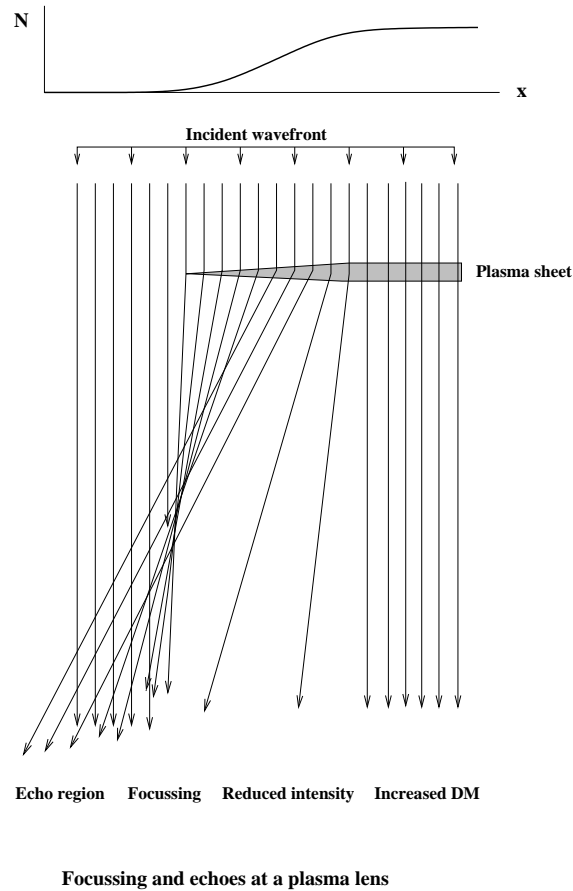}
\caption{Refraction at the tapered edge of a plasma cloud. The upper
curve sketches the variation of total electron content $N$ across the
edge of the cloud.}
\label{sketch}
\end{figure}

An ionised cloud acts as a diverging lens, since the refractive index
is less than one.  Clegg, Fey and Lazio (1998)\nocite{cfl98} analysed
the effect of a plasma lens with Gaussian profile on the apparent
brightness of a background source, with particular reference to the
effect of a cloud in the interstellar medium on a small diameter
extragalactic source; we show that their results are applicable to the
echo phenomenon.

Rays from a plane wavefront, passing through any centrally concentrated
lens, diverge and avoid a central disc, creating a shadow.  Figure
~\ref{sketch} shows the shadow behind the tapered edge of a planar
disc.  Here the wedge diverts rays outward, crossing the
undiverted rays and combining with them to increase the intensity.
Depending on the shape of the wedge and the distance of the observer,
the rays may combine to form a bright cusp.  Further away from the lens the
diverging rays give an increased intensity up to a maximum angle which
is the maximum refraction angle in the lens. In this region the source
is observed as double, with one image formed by rays traversing an
extra path. This is the region in which the echoes are observed.

This general picture applies to a cylindrical lens formed by a
filament and also to the edges of a plasma sheet.  The effect, as seen
on a screen, of refraction in a filament is an almost dark line with a
bright region on either side, with an extended halo.  The dark patch
is behind the region of the lens in which there is sufficient gradient
to divert the rays out of the line of sight.  If the source is pulsed,
the halo will show a delay which increases with angular distance.

The observed echoes conform well to this pattern.  As the lens
approaches the line of sight to the pulsar, the observer is traversing
the screen.  At the edge of the halo a faint echo is seen at maximum
delay.  As the delay decreases to zero the observer crosses the cusp,
with enhanced intensity. The main pulse then disappears behind the
lens (at day 740).

When the main pulse reappears 10 days later it builds up to normal
intensity, there is an increase in dispersion measure, and no clear
echo appears.  This indicates that the lens effect is at the edge of a
plasma sheet, in which there is a comparatively low gradient of
electron content.  A departing echo is observed starting 100 days
later, at a time of enhanced intensity, and coinciding with a steep
fall in dispersion measure. This is the other edge of the plasma
sheet, although it is less well defined as can be seen in Figure ~\ref{dispersion}b.

The scale on the screen can be found from the duration of the event,
given only the transverse velocity of the edge across the line of
sight.  Several examples of echoes have been observed at various
times, and all have similar durations and slopes.  We suggest that the
transverse velocity is almost entirely due to the proper motion of the
pulsar, which is observed to be 120 km s$^{-1}$ (Kaplan et al 2008)\nocite{kcga08}, with no significant contribution from the transverse velocity of the screen itself, and we
assume initially that the edge is traversed orthogonally.  The shadow
region extended for 10 days, i.e. $9\times 10^5$ seconds, giving a
scale of $1.0\times 10^{11}$ m.

The significant parameters of the lens are its integrated electron
content $N$ along a line of sight, and its gradient ${\rm d}N/{\rm
d}x$ across the line of sight.  The overall increase of DM is 0.10
cm$^{-3}$ pc, giving a maximum electron content $N_{\rm max}=3\times
10^{21}$m$^{-2}$. Averaged over the shadow region, the lateral
gradient ${\rm d}N/{\rm d}x$ was $3\times 10^{10}$ m$^{-3}$. Since
the echo was observed over a range of angles we assume that the cloud contains
structure with larger gradients which gave rise to the echoes at
the maximum delay.  We explore this by finding the angular deviation
corresponding to this averaged overall gradient, and comparing with an
independent estimate from the geometry of the observed delays.

The angular deviation in a gradient ${\rm d}N/{\rm d}x$ is found from
the phase change in traversing a plasma with total content $N$.  The
phase advance is $\lambda r_{\rm e} N$, where $r_{\rm e}$ is the
classical electron radius $e^2/(m_{\rm e}c^2) = 2.8 \times 10^{-15}$ m.  The
angular deviation $\theta =\lambda^2/(2\pi)r_{\rm e}{\rm d}N/{\rm
d}x$, where $\lambda=0.5$ m is the observing wavelength.  For the
gradient found above the average deviation $\theta_{\rm av}=3.2\times 10^{-6}$ rad.

\section{The delay geometry and the location of the plasma cloud}

 The echo was first observed 50 days before the delay reduced to
zero. At a velocity of 120 km s$^{-1}$ this is a lateral distance
$\Delta=5.2\times 10^{11}$ m.  Using the average deviation angle the
distance $R=\Delta/\theta_{\rm av}= 1.6\times 10^{17}$ m $\simeq 5$
pc, about 2.5 times the radius of the nebula; a larger local gradient
of electron content would give a larger deviation $\theta$ and
reasonably place the cloud within the nebula.

The echo delay $\delta$ is related to $R$ and $\theta$ by $\delta = {1
\over 2c}\theta^2R$.  For the observed maximum delay (5 milliseconds)
this yields a maximum deviation angle $\theta_{\rm max} = 4.3 \times
10^{-6}$, about 1.5 times the angle expected from the average gradient
of electron content.  The two independent estimates of $\theta$ are
seen to be reasonably consistent.

The smaller angles of deviation involved as the echo approaches zero
delay are derived from parts of the edge with lower gradient.  As
expected, the echo delay follows a parabolic curve, since it is
proportional to $\Delta^2$.

The distance $R$ derived above places the electron cloud outside the
nebula.  As noted already, it would be reduced if the cloud contained
gradients larger than the average: it would also be reduced if the
transverse velocity were reduced below 120 km s$^{-1}$, as would be
the case if the edge were traversed at a non-orthogonal angle.  Our analysis is
consistent with an electron cloud or filament located within the outer
part of the nebula, and with a gradient of electron content given by
the calculated average value.

\section{The non-dispersive delay.}

The element of delay, amounting to 1.2 ms, shown in Figure ~\ref{dispersion} 
to be
independent of frequency, must be related to ray paths in the electron
cloud.  A lateral gradient of electron content evidently exists in the
region traversed by the rays for some days after the event at day 747,
and this can account for a geometric delay similar to that observed in
the echoes.  For this delay to be non-dispersive, however, the
refracting region must be localised, so that rays at the three radio
frequencies traverse nearly the same geometric path. We suggest that
this path is close to the region where there gradient reduces to zero
and the profile of electron content becomes flat, as in the idealised
model of Figure ~\ref{sketch}.

\section{Comparable events in the Crab pulsar}

Almost 40 years of recordings exist of the radio pulses from the Crab
pulsar. Only three other events comparable to the 1997 event have been
recorded, in 1974, 1992 and 1994; echoes were observed in all three,
although in the 1974 event, which was the largest, there was
insufficient resolution of the echoes to allow a measurement of their
delays.  It was nevertheless remarked by Lyne and Thorne
(1975)\nocite{lt75} that the pulse profiles shown contained discrete
components, which we now interpret as echoes; and a re-examination of
their published profiles shows that all three components of the pulse
(precursor, main, and interpulse) were followed by echoes.  The pulse
intensity was observed to decrease to near zero in this event, as
occurred in 1997.  No other comparable event has been found in a close
scrutiny of more recent recordings (from 1997 to 2009).  The echo delays, and their rates of change, were similar to those in the 1997 event. In our interpretation, this allows us to assume that the edge of the plasma cloud is traversed approximately orthogonally, as in Section 4 above. 

Echoes of giant pulses from the Crab pulsar, with delays of 50 - 100
$\mu$s and durations of hours to some days have been observed by
Crossley et al (2007)\nocite{ceh07}, and interpreted using a
geometrical analysis similar to ours.  These echoes were attributed to
a plasma cloud closer to the pulsar, and probably within the diffuse
synchrotron nebula. Remarkably the scale of this structure and that of
the filament whose effects we have observed are closely similar, despite the different
locations in the inner and outer parts of the nebula; both
are around one astronomical unit.

Bhat et al (2008)\nocite{btk08} attributed slowly varying scattering of the Crab pulsar observed at 1400 MHz to small-scale filamentary structure in the nebula.   Kuzmin et al. (2008)\nocite{kljs08} found a close correlation between
the incidence of scattering at 100 MHz and changes in dispersion
measure of the Crab pulsar, on a time scale of some months. Further
observations at Jodrell Bank, and earlier work by Rankin and Counselman
(1973)\nocite{rc73} and by Isaacman and Rankin (1977)\nocite{ir77},
suggest that these variations are continuous, and not related to the
discrete events under discussion.  Kuzmin et al. do however suggest
that the source of the variable component of the dispersion measure is
within the Crab nebula, and that it again has a linear scale of around
one astronomical unit.

\section{The nature of the plasma cloud}

In all three events recorded with sufficient resolution the delay is
observed to decrease to zero and subsequently increase in a near
parabolic form.  This is interpreted as the two edges of a filament
crossing the line of sight.  In the 1997 event there appears to be a
relatively flat profile of electron content across the cloud, but we
may interpret the phenomenon generally as a filament with diameter
around $3\times 10^{11}$ m with a total line-of-sight electron content
around $3\times 10^{21}$ m$^{-2}$.  The electron density would then be
of order $10^{10}$ m$^{-3}$, i.e. $10^4$ cm$^{-3}$.  Direct
observation of such filaments by their emission or absorption is a
rather remote possibility; milliarcsecond angular resolution would be
required, in contrast to the 100 milliarcsecond resolution of the HST
observations of filamentary structure in the outer parts of the nebula
(Hester et al 1995)\nocite{hmb+02}.

The rarity of these events suggests that the Crab Nebula is only
sparsely filled with such filaments. If a typical filament takes
several days to cross the line of sight, and the interval between such
events is several years, we can envisage a nebula containing only a
few thousand such filaments, spaced apart by some hundreds of their
diameters and located only in the outer regions.  The filaments bear no
obvious relation to the well-known visible filaments, and do not contribute
significantly to the overall mass of the Nebula.

We thank the referee for helpful suggestions on the presentation of this paper.
\bibliographystyle{mn2e}
\bibliography{journals,psrrefs,modrefs}

\end{document}